\documentclass[10pt,conference,letterpaper]{IEEEtran}
\IEEEoverridecommandlockouts
\AtBeginDocument{%
  \providecommand\BibTeX{{%
    \normalfont B\kern-0.5em{\scshape i\kern-0.25em b}\kern-0.8em\TeX}}}

\usepackage{cite}
\usepackage{algorithm}
\usepackage{algpseudocode}

\usepackage{graphicx}
\usepackage{caption}
\usepackage{subcaption}

\usepackage[hidelinks]{hyperref}

\usepackage{amsmath}

\begin{document}

\title{Graph-based Modeling and Simulation of Emergency Services Communication Systems\thanks{This work is in part sponsored by the National Security Agency under Grant
Number H98230-20-1-0314. Any opinions, findings, and conclusions or
recommendations expressed in this material are those of the author(s) and
do not necessarily reflect the views of the National Security Agency. This
work is in part an outcome of InterPARES Trust AI, an international
research partnership led by Drs. Luciana Duranti and Muhammad Abdul-Mageed,
University of British Columbia, and funded by the Social Sciences and
Humanities Research Council of Canada (SSHRC).}}

\author{\IEEEauthorblockN{Jardi Martinez Jordan and Michael Stiber}
\IEEEauthorblockA{\textit{Computing and Software Systems Division} \\
\textit{School of Science, Technology, Engineering, and Mathematics}\\
\textit{University of Washington Bothell}\\
Bothell, WA, USA 98011 \\
jardiamj@uw.edu, stiber@uw.edu}
}

\maketitle

\begin{abstract}

Emergency Services Communication Systems (ESCS) are evolving into
Internet Protocol based communication networks, promising
enhancements to their function, availability, and resilience.
This increase in complexity and cyber-attack surface demands
better understanding of these systems' breakdown dynamics 
under extreme circumstances. Existing ESCS research largely overlooks
simulation and the little work that exists focuses
primarily on cybersecurity threats and neglects critical factors
such as non-stationarity of call arrivals. This paper introduces
a robust, adaptable graph-based simulation framework and essential
mathematical models for ESCS simulation. The framework uses a
representation of ESCSes where each vertex is a
communicating finite-state machine that exchanges messages
along edges and whose behavior is governed by a discrete event queuing
model. Call arrival burstiness and its connection to
emergency incidents is modeled through a cluster point process.
Model applicability is demonstrated through simulations
of the Seattle Police Department ESCS. Ongoing work is developing
GPU implementation and exploring use in cybersecurity tabletop exercises.

\end{abstract}


\section{Introduction}

An Emergency Services Communication System (ESCS) encompasses  organizational, electronic, and virtual elements that
answer emergency calls and  coordinate  responses. Despite their pivotal role as  critical
infrastructure, these systems
can exhibit vulnerabilities to cyber-attack~\cite{mirsky-guri20,xue-roy21} and  extreme
circumstances~\cite{dearstyne_fdny_2007,simon_world_2001}.
As such, understanding when and how their proper functioning starts to collapse
is essential.

ESCSes are currently undergoing significant infrastructure updates. In the United States, the Next Generation 911 (NG911) initiative
is converting the Enhanced 911 (E911) system into an Internet Protocol (IP)-based
system~\cite{nena_911_core_services_committee_2021}.
This IP-based ESCS infrastructure is also being implemented in 
Ecuador~\cite{corral-de-witt_e-911_2018}, Europe~\cite{liberal_european_2017}, and
Asia~\cite{markakis_emynos_2017}. This promises
enhancements to the function, availability, and resilience of these systems by
improving  inter-connectivity between Public Safety Answering Points (PSAPs) and
emergency responders, enhancing call localization, and allowing ESCSes to
receive text and multimedia messages on top of traditional voice calls. However,
this change increases system complexity and 
the cyber-attack threat surface.

PSAPs tend to be the main focus of modeling
and simulation research because of their critical role of connecting the public
with  emergency responders. Since PSAPs are emergency
call centers, analytical models such as  Erlang
equations are often used to model their operation~\cite{nena2003}. However, the
simplicity of analytical models requires that one makes significant and
often unrealistic simplifying assumptions; such models can be unsuitable for  complex systems, such as ESCSes.

This paper  introduces an adaptable graph-based simulation
framework and mathematical models that provide a comprehensive solution for
ESCS simulation. This  includes
abstract models for the three discerning aspects of an ESCS: (1)
network structure, (2)  internal behavior of its main components, and (3)  interactions among these components.

We designed our simulation framework using our general-purpose simulator for graph-based systems,
Graphitti~\cite{graphitti_2023}. ESCS components
are depicted as a graph-based structure supporting Communicating
Finite State Machines (CFSMs). Each vertex in this graph represents
an FSM operating within a discrete event queuing model;  each
edge denotes a communication link. Currently a sequential implementation,
the architecture is designed to facilitate  GPU 
implementation.

\section{Related Work}

Previous research  on ESCSes has predominantly focused on data
analysis and  development of predictive models aimed at  identifying
 incidents and call patterns; simulation 
remains relatively limited and unexplored. Reference \cite{neusteter_911_2019} analyzed 35
papers delving into 911 data within policing contexts, revealing two main
methodological approaches. One approach relies on standardized metrics such as call
volume, types, and response time, offering a comprehensive yet less detailed
perspective. The second approach employs complex techniques
to model caller behavior, track call-type patterns over time, and identify
obstacles to prompt responses. Notably, none of the studies in this review
took advantage of simulation.

The little work that uses simulation 
primarily concentrates on  vulnerabilities  to
cyber-attacks such as DDoS~\cite{mirsky-guri20,xue-roy21}.
The only exception is  \cite{gustavsson_thesis2018},
who employed simulation  within Operation Management of
an Emergency Call Center. This research introduced a method for modeling
burst behavior in the call arrival process that, although different
than the cluster point process presented here, also leverages
the link between emergency calls and the incidents that trigger
them~\cite{gustavsson2018}. Reference \cite{gustavsson_thesis2018} studies
important aspects of ESCSes such as the non-stationarity of call arrivals,
stochastic call-taker behavior, and the  relationship between
the geographical distance between PSAPs and their service area and service time. However,
their simulations were tailored to a Swedish emergency call center provider,
requiring extensive rework  for other emergency call centers.
In addition, their work recognizes the importance of call abandonment but
does not include it in their models.

Reference \cite{mirsky-guri20} researched the
vulnerability of the 911 system to DDoS attacks by
anonymous unblockable calls originating from mobile phones. They conducted practical
assessments by deploying bots  within a
limited cellular network. They then simulated
the potential impact  of a DDoS assault on an E911 network,
finding that fewer than 6000 bots could disrupt emergency services throughout a
state for extended periods.
Their simulation employed a lumped discrete event simulation of
the busiest time of the day to mirror the potential overload of Selective Routers due
to  a botnet; however, it lacked depth in representing the comprehensive
behavior of 911 entities such as PSAPs and responders.

Reference \cite{xue-roy21} also studied  DDoS attacks, using a
cyber-physical queuing-network model of the 911 system. Their simulation included models for a time-varying Poisson process
for call arrival, an exponentially distributed service time, dispatching dynamics,  responder
travel time, and response time. However, their study omitted other important
processes, such as call abandonment and redial, and focused on a small area of
10x15 miles (Charlotte, NC) with no discussion of model scalability.

\section{Contributions}

Rather than implementing a one-off simulation, we have developed a first-of-its kind,
generalized framework for simulating graph-structured systems and  applied it to ESCS operation. 
Our framework specifies graphs using  GraphML, a standardized graph representation language. Call arrival and other stochastic processes are specified using Extensible Markup Language (XML). This
provides flexibility for simulating heterogeneous ESCS networks and adapting
diverse mathematical models for the stochastic processes.
Modeling  call arrivals
is especially complex  due to its multifaceted nature, including inter-day,
intra-day, and seasonal variability~\cite{gustavsson2018}. A novel
contribution of our mathematical models is the modeling of call arrivals
as the realization of a cluster point process, capitalizing on the
connection between emergency calls and the underlying emergency incidents.
Moreover, our work delves into aspects of ESCSes  frequently
overlooked by other models, such as call abandonment, redial, and 
 emergency personnel dispatch.

\section{The 911 Response Process}

The 911 response is a dynamic process with multiple players.
It  starts with the person in need, who makes the initial
call for help that connects them to a call-taker at the nearest
PSAP. The call-taker enters essential details into
a system that pinpoints the caller's location and categorizes the type of
emergency. From there, a dispatcher, who might be the
call-taker in some cases, sends the right emergency responders to
the scene of the incident. Once the responders are on-site, they provide the
necessary assistance. But it does not end there; the system keeps
gathering valuable data. Responders file reports, cross-check
their assessment with the call-taker's information, and
record response times~\cite{neusteter_911_2019}.

\subsection{Call Initiation}

Beyond connecting callers with the appropriate PSAP,
telephony service providers are mandated to transmit the caller's
location --- a process that varies in form and precision depending on the calling technology: wireline phones, wireless cellular devices, or
Voice over Internet Protocol (VoIP) services. For wirelines, this would typically be a physical address; for wireless, it could involve GPS in the handset or cell tower information, which could result in location inaccuracy of kilometers~\cite{barnes_911_2014}.

\subsection{Call Routing and Delivery}\label{sec:call-routing}

Call routing aims to link the caller with the appropriate PSAP.
Previously, the emergency call was directed
to the nearest PSAP based on the caller's location, retrieved from the
local telecommunications network. However, the evolution to NG911 is ushering in a revised routing mechanism reliant on Geographic
Information System (GIS) data~\cite{fcc_legal_2013}. NG911 relies on GIS
databases that contain the service boundaries of PSAPs and emergency
responders, leading to the redirection of calls to the PSAP serving the
caller's precise location.

The new routing framework is tailored to modern
mobile calls and internet calling services, integrating wireline and older
wireless services through a Gateway,
thus standardizing their protocol. Furthermore, a Location-to-Service
Translation (LoST) protocol server unifies location information, either
civic addresses or geographic coordinates, into a Uniform Resource Identifier (URI)
that is used to route the emergency call. Ultimately, all emergency calls
go through the same LoST protocol
and routing engine.

\subsection{Call Processing and Dispatch}

After an emergency call reaches a PSAP, call-takers follow a specific
 protocol to assess the emergency, determine necessary
services, and provide  information for responders~\cite{neusteter_911_2019}.
While this protocol might vary among PSAPs, the National Emergency Number Association (NENA) outlines minimum requirements, suggesting that call-takers
should gather critical details such as the incident's address or precise
location, a callback number, the nature of the call, the time it occurred,
any identified hazards, and the caller's identity~\cite{association_nena_2020}.

NENA also establishes standards for the time intervals between each step
of the  process.
According to these guidelines, NENA recommends that 90\% of calls
should be answered within 15 seconds and
95\% of calls  within 20 seconds.

The call processing phase culminates ideally in
dispatching an emergency response unit to the incident.
Dispatch responsibilities vary between jurisdictions, 
often involving specialized dispatchers within certain PSAPs.
Once essential
information is gathered, call-takers proceed to either dispatch the appropriate
response unit or transfer the call to a designated dispatcher. This dispatching
process involves critical decisions regarding the types and quantity
of response units required. All pertinent information must
be communicated and logged into a Computer-Aided Dispatch (CAD)
system that assists responders in swiftly identifying
the required type and speed of response~\cite{neusteter_911_2019}.

\subsection{Incident Response}

The primary goal of an ESCS is
 ensuring  prompt and appropriate response. The interval between the initial call  and  
a response unit arrival at the scene is known as the \textit{response time}.
This time frame is critical within an ESCS and serves as a key metric
for evaluating performance, as noted in the work by \cite{stevens_response_1980}.
Reference \cite{stratmann_dial_2016} supports this by linking a
decrease in 911 response times to a significant reduction in homicide
rates.

Once a response unit is dispatched, it remains
occupied for the combined duration of the response time, the time it
takes to travel to the site of the incident, and the period
dedicated to delivering necessary services at the emergency scene.
This delivery of emergency services marks the final stage of the emergency
response process.

\section{Methodology}

Our  framework is designed to simulate
the operations of large heterogeneous ESCS networks, and to easily
adapt diverse mathematical models for the various stochastic processes
involved. It can be used for studies of
cybersecurity threats, catastrophic events, and
regular ESCS operation.
We model three crucial aspects of an ESCS:
(1) system \emph{structure},
(2) its main components' internal \emph{behavior}, and
(3) the \emph{interactions} among these components.
These considerations are examined in the following subsections in order, from general to
specific.

\subsection{Structure: ESCSes as Graph-based Systems}\label{sec:escs_graph_based}

Initiating a 911 call sets in motion
a cascade of events that involve various layers of technology
and emergency personnel. ESCSes are complex heterogeneous
multi-network systems that have safety-critical, operational,
and regulatory concerns. For any model to be useful, however, it must be
simpler than the real system; thus, one must reduce it to the
core elements. In light of this,
the following three fundamental ESCS entities were identified
and characterized:
(1) Caller Regions (CRs),
(2) PSAPs, and
(3) First Responders.

Here, \textit{Caller Regions} denote the geographic areas from where calls
originate, \textit{PSAPs} are emergency call centers responsible for answering
emergency calls and dispatching first responders, and \textit{Responders}
represent emergency response centers (e.g. police
and fire stations). These components are arranged in a network that
underlies the GIS-based call routing and dispatching dynamics of an ESCS,
as described in \autoref{sec:call-routing}.

ESCSes are part of a family of complex
cyber-physical systems that are effectively
represented using graphs. Graphs are used to model network nodes
(\textit{vertices}) and their connections (\textit{edges}) in a
pairwise relationship. The particular model, in our framework, is a
\textit{directed graph}, where vertices denote the aforementioned
ESCS entities and edges represent the communication channels between them.

\subsection{Behavior: Discrete Event Queueing Model}\label{sec:queuing_model}

In essence, PSAPs are call centers specialized for emergencies.
Therefore, discrete event queuing models extensively used in the
management of call centers are suitable for modeling the processing
of emergency calls by PSAPs. Furthermore, the same concept can be
applied to emergency responder nodes, thus forming a multi-queuing model.

Conceptually, a call center contains $k$
trunk lines with up to the same number of workstations ($w \leq k$)
and agents ($n \leq w \leq k$). The number of trunk lines corresponds to the number
of simultaneous calls, so the number of agents plus the capacity
of the center's queue. One of three scenarios occurs when a call
arrives: (1) it is answered right away if there is an available agent,
(2) the call is placed in the queue (on hold) if there are no available agents, or
(3) the caller receives a busy signal if there are no trunks available.

In this model, we think of an agent as a resource that is occupied while a
caller is receiving assistance and immediately released once it has been
served. Calls are lost due to blocking when all trunks are busy or
when a caller abandons the queue due to impatience, possibly redialing
soon after. Consequently, arriving calls come from either those who
made an initial call, those who got a busy signal, or those who
abandoned the queue after waiting.

Stochastic processes such as call arrivals, service time, and customer
impatience ---  call abandonments --- must be modeled through
random variables, drawn from  suitable probability distributions.
Existing research in this area provides  candidates but their
goodness of fit must be evaluated based on data obtained from the
real system~\cite{law_simulation_2000}.
The mathematical model for the call arrival process
is discussed, in detail, in \autoref{sec:call-arrival}; the
 details of other stochastic processes  are
presented in \autoref{sec:stochastic-processes}.

\subsection{Interactions: Communicating Finite State Machines}\label{sec:cfsm}

\begin{figure*}%[!htb]
    \centering
    \includegraphics[width=0.75\linewidth]{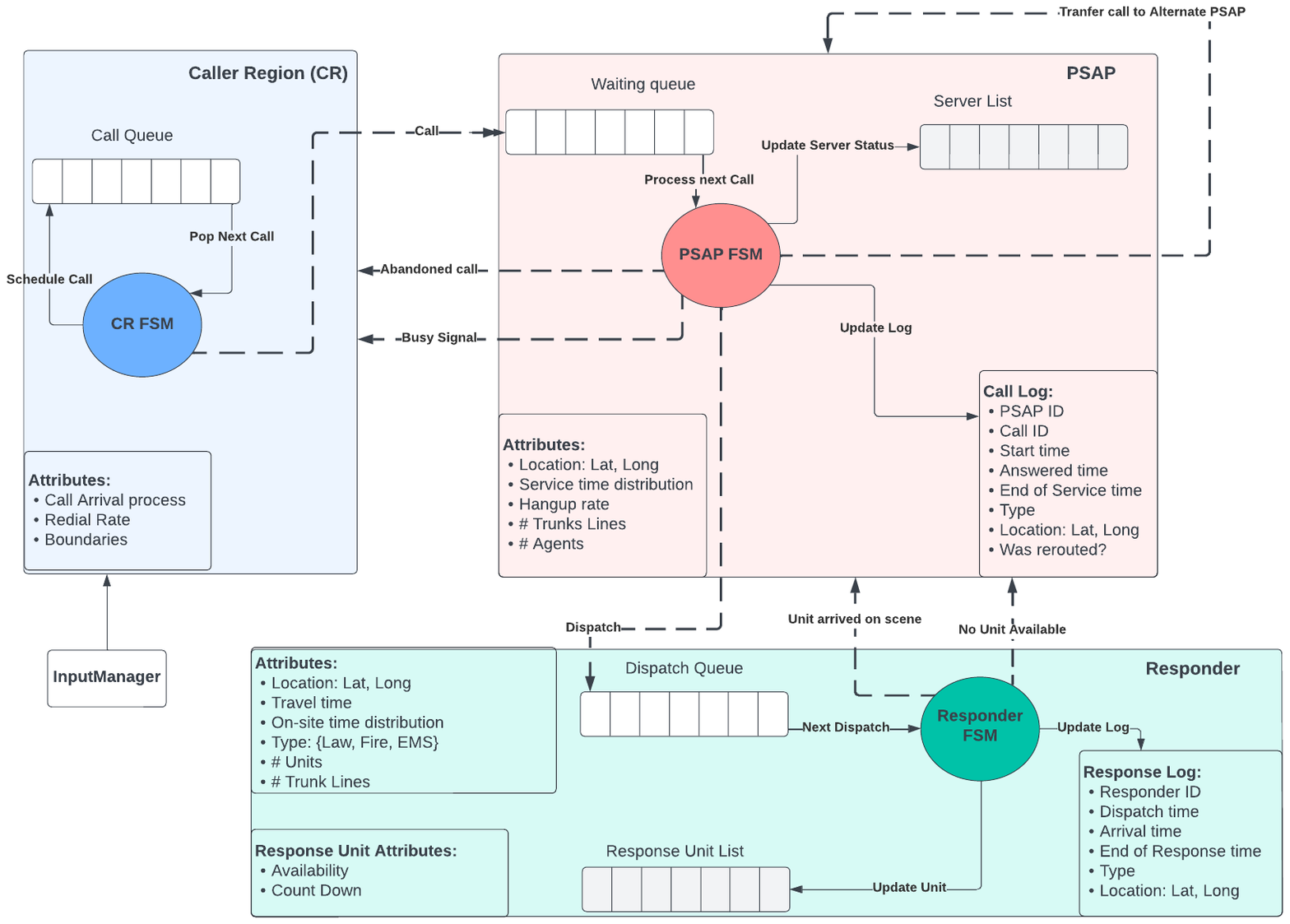}
    \caption[911 Communication Finite State Machine]
            {911 Communicating Finite State Machine model of an ESCS.}
    \label{fig:cfsm}
\end{figure*}

Communicating Finite State Machines (CFSMs) were researched by
\cite{brand_communicating_1983} in the domain of distributed
computing. They employed a common representation of communicating
processes, where each process functions as a finite-state machine
and inter-process communication occurs through reliable full-duplex
first-in first-out (FIFO) channels. CFSMs were chosen for modeling
the interactions between ESCS entities because they offer several advantages, 
such as the ability to analyze the execution of each process separately,
rather than considering the system as a whole, thus partitioning 
complex behavior into manageable sub-problems. Notably, this model abstraction
aligns with the potential for a future parallel GPU implementation, which Graphitti is designed to facilitate.

A simulation begins with an initial configuration defined by
specific parameters and proceeds for a predetermined number of
time steps. The values of these parameters and simulation
variables at each time step collectively constitute
the ``system state.'' For an ESCS graph, the system state
fully represents every vertex and edge
it contains. During each time step, ESCS entities
receive inputs and undergo state transitions, potentially
resulting in the transmission of outputs to other vertices.

\autoref{fig:cfsm} illustrates how ESCS entities, modeled
as vertices, receive messages through
incoming edges and transmit messages via outgoing edges.
Each vertex is depicted as a Finite State Machine (FSM),
with state transitions determined by events arising within a 
discrete event queuing model, such as call initiation, call answer,
unit dispatch, call termination, and end-of-emergency response.
Consequently, each vertex contains a First-In-First-Out (FIFO) queue that
dictates event processing order and a
set of attributes, some modeled as stochastic processes with
probability distributions.
The PSAP and responder vertices maintain lists of
resources (servers and responder units, respectively)
and record metrics for serviced calls
and emergency responses.

\subsection{Call Arrival as the Realization of a Spatiotemporal Cluster Point Process}\label{sec:call-arrival}

A \emph{spatiotemporal cluster point process} is a random process that captures
the clustering of events in space and time, characterized by three key elements:
(1) a primary process, (2) a secondary process, and
(3) a systematic aggregation of secondary points.
In simpler terms, a cluster point process aggregates clusters of events
associated with the points of a primary process, often called the ``parent
process''~\cite{serfozo_1990}.

Cluster point processes, like  Neyman-Scott~\cite{neyman_scott_1958},
excel in modeling natural phenomena such as cosmology~\cite{neyman_scott_1958},
rainfall analysis~\cite{Cowpertwait_2002}, seismic events~\cite{Anwar_2022}, and neural
spike trains~\cite{gomez-etal05}, highlighting their accuracy
in capturing complex real-world phenomena. In this work, we use
a spatiotemporal cluster point process for modeling the relationship between
emergency calls and the emergency incidents that generate them. The
primary process denotes emergency incidents, whereas the secondary process
represents resulting calls, effectively capturing
the spatial and temporal clustering of calls near emergencies.

\subsubsection{The primary process}

Any point process that effectively captures emergency events' temporal and spatial
characteristics can serve as the primary process. The
distributions of points in time and space operate independently,
accommodating stochastic and deterministic occurrences, and regular and
irregular patterns. Here, we use a Poisson distribution to model the
primary process in time, with events assumed to be uniformly
distributed in space.

\subsubsection{The Secondary Process}

Emergency calls are simulated as a Poisson point process inside a circle, centered
at the corresponding primary event. For each cluster, we
model the number of calls, the calls' arrival times, and the
location of each call.

Each cluster's circle is defined by a radius $r>0$ and an intensity $i>0$.
The intensity indicates the number of points per unit area; thus the number of
secondary points ($n$) is
\begin{equation}
n = \pi \cdot r^2 \cdot i
\end{equation}

Our model captures the varying magnitudes of emergency events through the use of
\emph{prototypes} that define
 radius and intensity. In practice, the prototypes' values and rate of
occurrence are determined by the incidents and call patterns
observed at the PSAPs being studied. The parameters of the secondary process are then drawn
from a normal distribution that aligns with the selected prototype.
Given a prototype radius' mean  ($\mu_r$) and
standard deviation ($\sigma_r$),  the cluster radius is defined as
\begin{equation}
R \sim \mathcal{N}(\mu_r, \sigma_r^2)
\end{equation}
where $R$ follows a normal distribution.
Likewise, from a prototype intensity's mean  ($\mu_i$) and standard deviation ($\sigma_i$),
a normally distributed intensity is defined as
\begin{equation}
I \sim \mathcal{N}(\mu_i, \sigma_i^2)
\end{equation}

The calls' arrival times are computed as follows: let $T$ be a
random variable representing the calls' inter-arrival interval. Then,
$T$ follows an exponential distribution
\begin{equation}
    T \sim \text{Exp}(\sigma_{t})
\end{equation}
with rate parameter $\sigma_{t}$.
Subsequently, the cumulative sum of $T$ (\(t_1, t_1 + t_2, ..., t_1+t_2+t_3+...+t_n\))
is added to the timestamp of the emergency incident.

% \begin{algorithm}
% \caption{Generate secondary events}
% \label{algo:sec_process}
% \begin{algorithmic}[1]
% \ForAll{prim\_event \textbf{in} prim\_events}
%     \State Select $\mu_r, \sigma_r, \mu_i$, and $\sigma_i$ from a prototype 
%     \State Sample $R$ from a normal distribution $\mathcal{N}(\mu_r, \sigma_r^2)$
%     \State Sample $I$ from a normal distribution $\mathcal{N}(\mu_i, \sigma_i^2)$
%     \State Calculate $n = \pi \cdot R^2 \cdot I$
%     \ForAll{$n$ secondary points}
%         \State Sample $T$ from exponential distribution (rate $\sigma_t$)
%         \State Sample $U$ from uniform distribution [0,1]
%         \State Sample $V$ from uniform distribution [0,1]
%         \State Calculate $(\rho,\theta) = (r\sqrt{U}, 2\pi \cdot V)$
%         \State Calculate $(x, y) = (\rho \cdot \cos(\theta), \rho \cdot \sin(\theta))$
%         \State Calculate $t = \text{prim\_event}(t) + cum\_sum(T)$
%     \EndFor
% \EndFor
% \State \textbf{return} $sec\_events(t, x, y)$
% \end{algorithmic}
% \end{algorithm}

For the spatial domain, the emergency calls are scattered within the
circle using polar coordinates ($\theta$, $\rho$), drawn from a uniform distribution.
The distribution of angle values ($\theta$) is proportional to $2\pi$, whereas
the area of a circle is proportional to the square of its radius ($\mathit{Area} = \pi \cdot r^2$).
Therefore, if $U$ and $V$ are two independent uniform random variables on the
interval (0, 1), the polar coordinates of points uniformly located on a circle of
radius $r$ are given by
\begin{equation}
(\rho, \theta) = (r \cdot \sqrt{U}, 2\pi \cdot V)
\end{equation}

\subsection{Modeling  Other Stochastic Processes}\label{sec:stochastic-processes}

\subsubsection{Call Abandonment}\label{sec:abandonment}

A customer who calls when all servers are busy is placed in a waiting queue
(see \autoref{sec:queuing_model}); those who run out of patience
before their call gets answered, hang up. This process is known as \emph{abandonment}.
The idea of customers being prepared to wait for a specific duration is used
in the Palm/Erlang-A model to incorporate call abandonment into the Erlang-C
equation. In the Erlang-A model, each call arrival is
associated with an exponentially distributed \textit{patience time} with
mean $\theta^{-1}$ and gets an \textit{offered waiting time} in the
queue. If the waiting time exceeds the patience time, the
call is considered abandoned~\cite{mandelbaum2007}.

The importance of incorporating call abandonment into a model is illustrated
by  \cite{gans2003telephone} and \cite{mandelbaum2007}.
Reference \cite{gans2003telephone} used the Erlang A queuing model,
demonstrating that even a small fraction of abandoned calls during heavy
traffic significantly impacts system performance.
Similarly, \cite{mandelbaum2007}
showed that in a heavily loaded call center, 3.1\% abandonment reduces
the \textit{average speed of answer} from 8.8 seconds to 3.7 seconds,
because abandonment effectively decreases the workload
precisely when it is most needed.

We adopt an exponentially distributed patience time. However,  
the abandonment rate $\theta$ is not directly observable because
patience time is only measurable for customers who abandon the queue.
For those who receive service, the waiting time is only a lower bound
of their patience time. Here, we implement the method suggested
by \cite{mandelbaum2007}, based on the
relationship between the average wait in the queue ($E[W]$) and the fraction
of customers that abandon it ($P\{\mathit{Ab}\}$), for estimating the abandonment rate 
$\theta$ as
\begin{equation}
\theta = \frac{P\{Ab\}}{E[W]} = \frac{\text{Abandonment Fraction}}{\text{Average Wait}}
\end{equation}

\subsubsection{Redial}

If all call takers and trunks are busy when a customer
initiates a call, they receive a busy signal and the call is dropped.
NENA determines that 85\% of customers whose calls drop
immediately redial. In our model, the redialing
decision is represented by a discrete random variable ($\mathit{Rd}$) that follows
a Bernoulli distribution with  probability of success of $P = 0.85$.
The redialing probability can be fitted to a given region from call data
that uniquely identifies the callers. However, obtaining call data with
unique identifiers from the 911 system was unfeasible for our work
due to technical obstacles, privacy considerations, and legal
compliance issues.

\subsubsection{Service Time}

Service time
represents the interval between the initial call answer and the dispatch
of a first responder to the emergency incident. Here, 
service time is assumed to follow an exponential distribution and is
predetermined at the outset of the simulation.

\subsubsection{Responder Dispatch and Response Time}\label{sec:response-dispatch}

Dispatching  responders starts with selecting  the most suitable one
for a given emergency event. This involves  consideration of
responder availability, expertise, and proximity to the event.
In our modeling framework, responder units
are dispatched from different stations, with priority given to the
unit stationed closest to the emergency incident.

To enhance simulation realism, emergency calls, and therefore responses, are
categorized into one of three types:
(1) Law: requiring a police response,
(2) EMS:  requiring an Emergency Medical Service response, or
(3) Fire:  requiring a fire unit response.

The time a response unit remains occupied includes both the driving time to the incident
location and the on-site time dedicated to addressing the emergency.
Driving time is determined by calculating the time it takes for a
responder to cover the Euclidean distance between the dispatching center
and the incident location. The duration of on-site time is assumed
to be exponentially distributed, with
an average on-scene time of 20 minutes considered reasonable based
on prior research studies~\cite{david_retention_2009,vincent-lambert_views_2018}.

\begin{figure}[!bth]
    \centering
    \includegraphics[width=\linewidth]{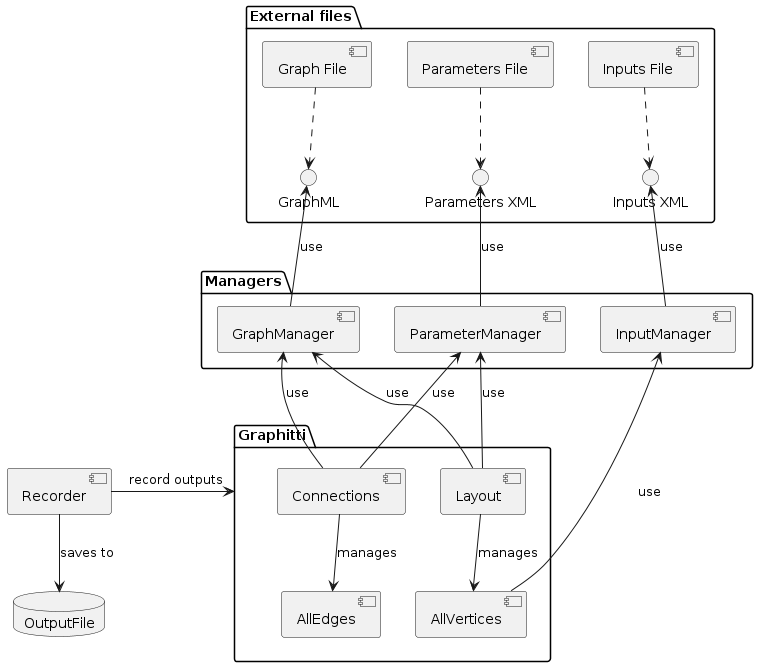}
    \caption[New Component Diagram]
            {Graphitti's Component Diagram.}\label{fig:component_diagram}
\end{figure}

\subsection{Implementation Within the Graphitti Simulator}

These simulations 
were implemented using Graphitti, our in-house graph-based simulator~\cite{graphitti_2023}.
Graphitti is designed to simulate discrete and continuous graph-based
systems, and facilitate the migration and
validation of large-scale (tens of thousands of vertices, millions of
edges) and long-duration (billions of time steps) simulations to GPUs.

\autoref{fig:component_diagram} provides a component diagram for our simulation  within Graphitti's architecture, where the ESCS graph,
simulation parameters, and input modeling are fed to the simulator through
external files, with graphs and call arrivals created by external Python scripts.

The representation of a graph within Graphitti is contained in four components:
Connections, Layout, AllEdges, and AllVertices. The \textit{Layout} component
manages the location and parameters of the vertices contained in
\textit{AllVertices};  \textit{Connections} manages the properties
of the edges contained in \textit{AlllEdges}.
\textit{Connections} and \textit{Layout}, in addition to using the
parameters from the \textit{ParameterManager} component, rely on the graph contained within
\textit{GraphManager} to generate Graphitti's internal representation of the
network. The  \textit{InputManager} loads the list of events
into internal queues, one per vertex, to be retrieved by the respective
vertex during the simulation. This design enables  handling  a wide range
of graph structures and input sources,  real-world 
or synthetically generated. As a result, one can conduct experiments by adjusting
 network structure or  input arrival rate without making changes to the
simulator's  code.

\subsection{Vertex Communication and State Transition}

%% Rewritten for brevity and removal of flow diagram
The CFSM abstraction allows us to consider the execution sequence of each
vertex independently, with interactions occurring through message
exchanges across connecting edges. This leads to a system with two
distinct phases: (1) a \emph{Communication Phase}, where vertices receive
messages, and (2) a \emph{State Transition Phase}, where vertices undergo
state transitions due to internal or external events.

During the \emph{Communication Phase}, vertices pull messages from
their incoming edges such as emergency calls, dispatch instructions,
and availability status as illustrated in \autoref{fig:cfsm}. Although
our current implementation is a sequential loop over all incoming edges,
this framework was purposely designed for future GPU parallelization.
This ``pulling strategy'' prevents multiple processes from simultaneously
accessing and modifying data structures, eliminating the need for locking
and enabling the use of a parallel reduction algorithm for efficiently managing numerous incoming edges. Each vertex stores the received messages
in a waiting queue if available space exists, otherwise, the message is
dropped and subsequently handled by the source vertex.

During the \emph{State Transition Phase}, vertices respond to internal
and external events, potentially undergoing state transitions.
As presented in \autoref{sec:queuing_model}, the state
transitions in each ESCS entity follow a discrete-event queuing
model, further discussed in this section.

\subsubsection{Caller Region State Transition}

%% Rewritten for brevity and removal of flow diagram
A Caller Region defines the geographical area where calls originate,
modeled as a cluster point process. Graphitti simulations are organized
into epochs, each comprising multiple time steps. As illustrated in
\autoref{fig:cfsm}, each caller region contains a call queue that,
in between epochs, is loaded with the calls scheduled for the next
epoch. During each time step, the region performs two actions:
(1) checks for dropped calls and initiates redialing if necessary, and
(2) routes the next call to the appropriate PSAP via an outgoing
edge. The current implementation ensures that only one call is
processed per Caller Region per time step, achieved by subdividing
a PSAP's service zone or adjusting the time step duration.

\subsubsection{PSAP State Transition}

%% Rewritten for brevity and removal of flow diagram
The primary role of a PSAP is to handle emergency calls from its
Caller Regions and coordinate the dispatch of appropriate emergency
responders. Each PSAP contains a waiting queue and a list of
servers as seen in \autoref{fig:cfsm}. The waiting queue holds
the calls not yet answered by a ``server'', responsible for
call handling. At each time step, a PSAP undertakes two key actions:
(1) managing servers that have concluded their service for a call, and
(2) assigning new calls to available servers.

\begin{figure*}%[!htb]
\centering
    \begin{subfigure}{.49\textwidth}
      \centering
      \includegraphics[scale=0.5]{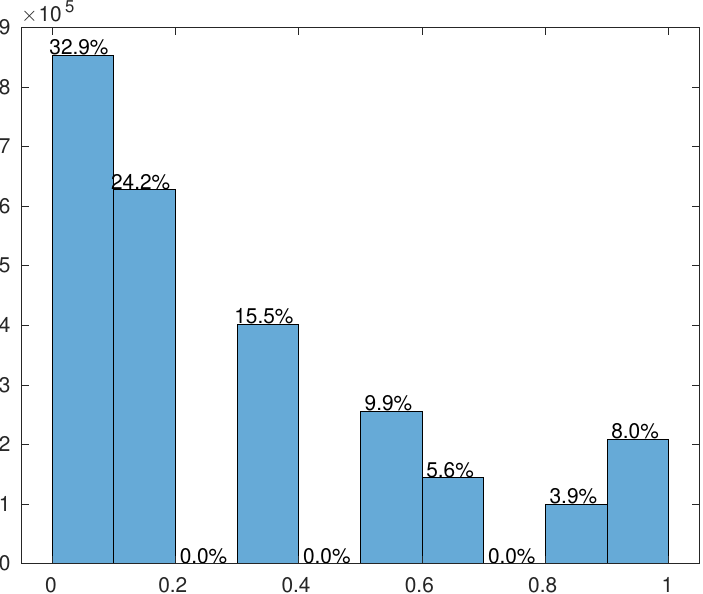}
      \caption{Arrival rate = 45.6 calls/hr}
    \end{subfigure}
    \begin{subfigure}{.49\textwidth}
       \centering
       \includegraphics[scale=0.5]{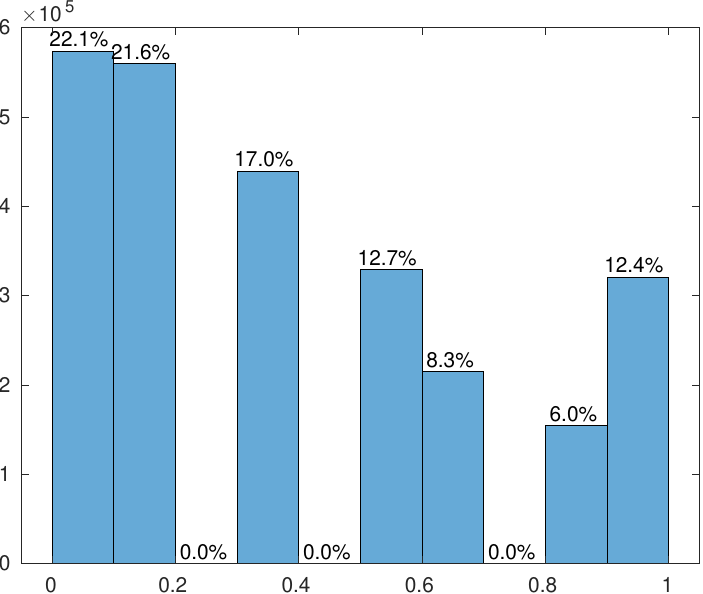}
       \caption{Arrival rate = 63 calls/hr}
    \end{subfigure}
    \begin{subfigure}{.49\textwidth}
       \centering
       \includegraphics[scale=0.5]{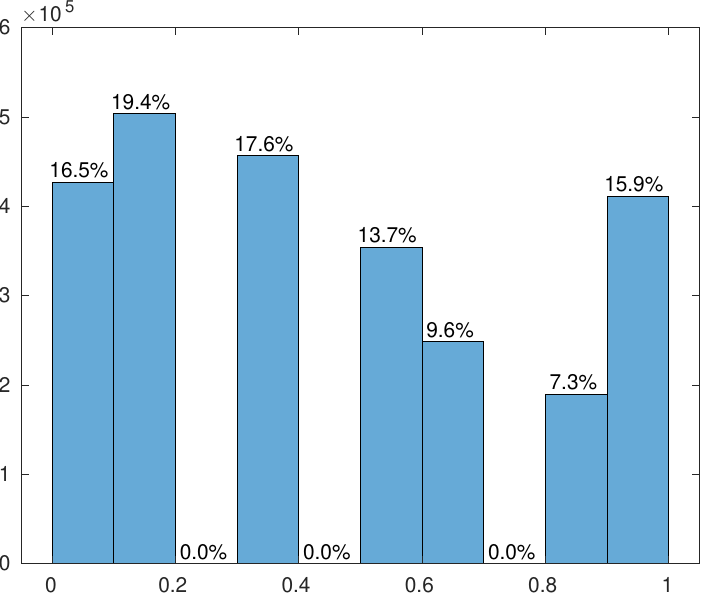}
       \caption{Arrival rate = 76 calls/hr}
    \end{subfigure}
    \begin{subfigure}{.49\textwidth}
    \centering
       \includegraphics[scale=0.5]{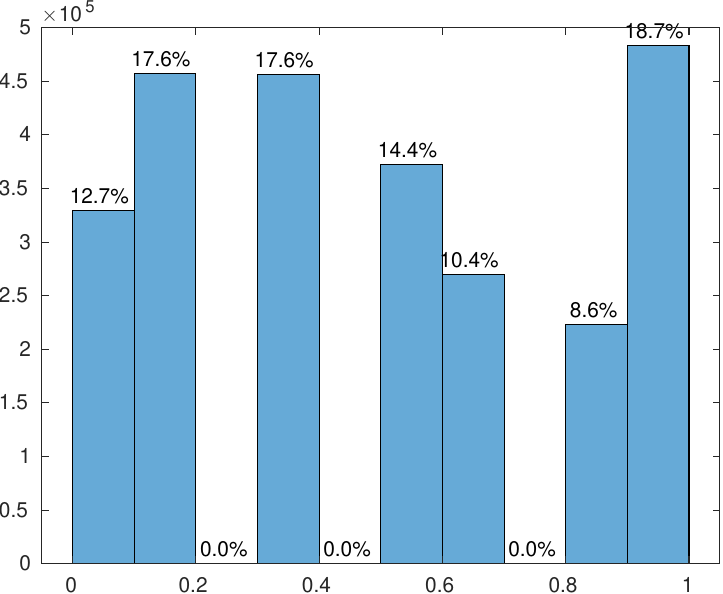}
       \caption{Arrival rate = 84.2 calls/hr}
    \end{subfigure}
    \caption{Call arrival rate impact on system utilization}\label{fig:utilization-histogram}
\end{figure*}

After completing a call, a server becomes available for new ones.
However, before taking on new calls, servers fulfill dispatch
responsibilities such as sending dispatch messages to the nearest
suitable responder via an outgoing edge and
record essential call metrics, such as start time, response
time, conclusion time, and instances of call abandonment.

Available servers are processed iteratively for call assignments
until no more servers are available or calls remain in the
waiting queue. Upon receiving a call assignment, a server is occupied for the
duration of the call. Additionally, call abandonments are managed within
this process as described in \autoref{sec:abandonment}. If a caller's wait
time exceeds their patience threshold --- the duration they are
willing to wait in the queue --- the call is classified as abandoned.

\subsubsection{Emergency Responder State Transition}

%% Rewritten for brevity and removal of flow diagram
A ``Responder'' vertex represents an emergency response center.
As illustrated in \autoref{fig:cfsm}, responder vertices operate
in a discrete event queuing model similar to PSAPs. Response
units address emergency events from a dispatch queue in two
phases. First, units completing an emergency response document
their metrics and return as available units. Then, new emergency
incidents are allocated to available units until all units are
busy or no incidents remain in the queue. Once an incident
has been assigned, a unit remains occupied for the duration
of travel to the incident location plus the time spent on-scene.

\section{Illustrative Application}

To demonstrate the applicability of our graph-based simulation framework,
we developed a graph representation of the ESCS network of
King County, Washington, extracted from a Geographic Information
System (GIS) dataset (provided by the Washington State
911 Coordination Office) comprised of five layers that map the
boundaries of PSAPs, police, fire, and EMS services. We  focused on the Seattle Police
Department (PD) composed of 1 PSAP, 34 Fire Stations (also providing
EMS services), and 5 Police stations. The Seattle PD PSAP, now known
as Community Assisted Response and Engagement (CARE), is the largest
of the 12 PSAPs in King County, serving about 10\% of all 911 calls.

A month-long dataset composed of 41,217 calls with a call rate of
57.25 calls/hr was used to parameterize the discrete event queuing
model for CARE. The dataset revealed a maximum hourly call rate
of 137, an abandonment rate of 9.42\%, and call durations ranging
from 4 to 2016 seconds (204 seconds average). Using Erlang
equations, estimations suggest a need for 6 call takers,
considering a 10-second caller waiting tolerance (Erlang C), and
16 trunk lines to manage a 1\% probability of blocking during
peak times with an additional 10 seconds of service time for
post-processing.

We conducted simulations using one month of synthetic call data
generated through the cluster point process algorithm discussed
in \autoref{sec:call-arrival}. These simulations used call arrival rates ranging from 45.6 to 84.2 calls/hr in
10\% increments. Other parameters were kept consistent throughout
the simulations. The redial probability used was 85\% with an
average patience time of 49.36 seconds (see \autoref{sec:abandonment}).
Additionally, the average on-scene duration was set at 20 minutes
(see \autoref{sec:response-dispatch}), whereas the minimum and
average service times were set to 4 and 204 seconds, respectively.
Simulations were run on a computer with a 14-core Intel i7-12700H 2.3 GHz processor, requiring 47 seconds for a one-month simulation of this single-PSAP network.

The selected metrics for analyzing simulation results were customer
waiting time and system utilization. System utilization is defined as
the fraction of  call-takers or responder units busy at
any given time. Utilization histograms in
\autoref{fig:utilization-histogram} are presented for the different arrival rates.
At 45.6 calls/hr, the system
surpasses 80\% utilization merely 11.9\% of the time, coinciding with an
average waiting time of 6.8 seconds. As the arrival rate grows to 84.2
calls/hr, utilization above 80\% increases to  27\%; the average waiting
time increases moderately to 9.4 seconds. However, at this threshold the
abandonment rate triples from 9 to 27.3\%.

\section{Conclusion}

In this work, we presented a detailed architectural framework
and mathematical models for simulating
 Emergency Services Communication Systems, designed to
easily accommodate diverse ESCS networks and their stochastic processes.
In addition, we introduced a novel approach for modeling  call
arrival burstiness as the realization of a cluster point process,
capitalizing on the relationship between emergency calls and
underlying emergency events. We then used our framework to implement
one-month-long simulations of the Seattle Police Department ESCS
with varying call arrival rates. 

We are currently implementing a GPU  version of
our models within Graphitti to improve  performance for long durations and  large
ESCS networks. We also plan to evaluate the use of our
simulator to support cybersecurity tabletop exercises.

\section*{Acknowledgment}

We extend our appreciation to additional contributors.
Scott Sotobeer offered invaluable insights into the 
workings of ESCSes and their
associated policies. Bennett Ye and Jacob White developed tools for ESCS graph visualization and editing,
and Divya Kamath and Xiang Li made substantial architectural
improvements to Graphitti.

\bibliographystyle{ieeetr}
\bibliography{escs}

% \appendix

% \section{Illustrative Application}




\end{document}